\documentclass[num-refs]{wiley-article}

\usepackage{color}
\usepackage{ulem}

\newcommand{\add}[1]{\textcolor{blue}{#1}}
\newcommand{\delete}[1]{\textcolor{red}{\sout{#1}}}
\newcommand{\edit}[2]{\textcolor{red}{\sout{#1}} \textcolor{blue}{#2}}
\newcommand{\mnote}[1]{\marginpar{\textcolor{green}{\textbf{#1}}}}

\usepackage{siunitx}

\papertype{Original Article}

\title{On the thermodynamic invariance of fine-grain and coarse-grain fluid models}

\author[1]{Thomas Dubos}

\affil[1]{LMD/IPSL, École Polytechnique/IP Paris}

\corraddress{T. Dubos, LMD/IPSL, École Polytechnique/IP Paris, 91120 Palaiseau, France}
\corremail{dubos@lmd.ipsl.fr}

\fundinginfo{ This work is part of the AWACA project that has received funding from
the European Research Council (ERC) under the European Union’s Horizon
2020 research and innovation programme (Grant agreement No. 951596) }

\runningauthor{T. Dubos}

\begin{document}

\maketitle

\begin{abstract}
In models of oceanic and atmospheric flows, thermodynamic functions and
conservative variables may be defined up to a certain
degree of arbitrariness, in the sense that, for a given set of observable quantities such
as pressure and temperature, the predictions of the
model are insensitive to, e.g. some reference enthalpies,
entropies or pressures.

Since the compressible Navier-Stokes-Fourier model, regarded as a
``mother'' fine-grain model, is invariant with respect to arbitrary changes
in reference enthalpies and entropies, restricted only by phase change,
any coarse-grain model obtained, even conceptually, from it must be
invariant at least to the same extent. Upon examination, it is found
that the dependence of usual conservative variables to a reference
pressure propagates to their fluxes and gradients, and to down-gradient
closures based on them.

Conversely, closure relationships between adequately defined ``reduced''
gradients and fluxes of enthalpy and entropy, are guaranteed
to lead to invariant models, even with multiple turbulent diffusivities and cross-diffusivities. More work is required to address the invariance of
more sophisticated closures, especially shallow and deep convective
closures.

\end{abstract}

\section{Introduction}

Thermodynamics are at the heart of fluid dynamics, especially of geophysical
fluid dynamics. The pressure force is of thermodynamic origin, latent
heat release drives moist convection, molecular conduction and diffusion,
despite their acting at small scales, are ultimately responsible for
irreversible transfer of heat and mass by turbulence. Nevertheless,
the theoretical treatment of thermodynamics in geophysical models
has often been limited to special cases such as dry air modelled as
an ideal perfect gas \citep{ogura_scale_1962,lipps_scale_1982,mellor_development_1982,durran_improving_1989}
or a mixture of species with temperature-independent heat capacities
\citep{lauritzen_ncar_2018}. With a few exceptions \citep{ooyama_thermodynamic_1990,bannon_hamiltonian_2003},
it is only recently that a significant effort has been put into accomodating
general thermodynamics of single-component or multicomponent fluids
\citep{pauluis_thermodynamic_2008,feistel_thermodynamic_2010,vasil_energy_2013,tort_usual_2014,tailleux_local_2018}
Thermodynamic potentials are the tools that allow for such a general
treatment of complex equations of state \citep{hauf_entropy_1987,feistel_thermodynamic_2010,thuburn_use_2017,eldred_thermodynamically_2022}.

Useful geophysical models result, at least conceptually, from a sequence
of approximations and coarse-graining steps, starting from a ``mother''
model such as the compressible Navier-Stokes-Fourier (NSF) equations.
The problem of constructing consistent approximations for the reversible,
adiabatic part of models is now essentially solved. Variational and
Hamiltonian methods \citep{eckart_variation_1960,salmon_practical_1983,holm_eulerpoincare_1998,morrison_hamiltonian_1998}
have been shown to be powerful tools to systematically obtain approximations
of geometric, hydrostatic or sound-proof type while guaranteeing conservation
laws \citep{vasil_energy_2013,tort_usual_2014,dubos_equations_2014}.
These approximations are relevant for both fine-grain (non-hydrostatic)
models and coarse-grain models.

The situation regarding coarse-graining is quite different. 
Coarse-graining incurs a loss of information that, together with nonlinear interactions
leads to the need to represent irreversible processes,
including those that have been averaged out, with closures. Building
closures remains a formidable challenge, and any guidance based on
theoretical considerations would be useful \citep{shaw_theoretical_2009}.
However for unaveraged, but approximated equations, it is only recently
that viscosity, conduction, diffusion and the associated entropy production
have received a comprehensive treatment \citep{pauluis_thermodynamic_2008,klein_thermodynamic_2012,eldred_thermodynamically_2021}.
Thus it is perhaps not surprising that such guidance is currently
very limited, even in terms of what energy conservation principle
closures should obey \citep{lauritzen_reconciling_2022}. Especially,
to which extent thermodynamical constraints such as the second law
of thermodynamics restrict the space of permissible closures remains
an open question \citep{goody_sources_2000,akmaev_energetics_2008,gassmann_entropy_2018,gassmann_entropy_2019}.
The present work is a step towards the goal of establishing such constraints
on firm ground.

There are several procedures to obtain a coarse-grain model from a
fine-grain model. Large-eddy simulations employ an explicit spatial
filtering procedure \citep{sagaut_large_2006}. Single-column parameterizations
result rather, at least conceptually, from Reynolds or ensemble averaging
\citep{mellor_development_1982}. Other averaging procedures, such
as multiple-scale expansions \citep{shaw_theoretical_2009}, isopycnal
averaging \citep{young_exact_2012} and conditional filtering \citep{thuburn_two-fluid_2022},
have also been proposed. To the author's knowledge, the coarse-graining
process that would lead from e.g. the NSF equations to a hydrostatic
general circulation model with closures for turbulence, shallow and
deep convection, is not known precisely. Thus, the present work focuses
on constraints that apply to coarse-grained models independently from
the details of the coarse-graining process, specifically invariance
properties. Indeed, if a fine-grain model possesses an invariance
property, and there is no good reason why the coarse-graining process
would break it, the resulting coarse-grain model must enjoy at least
the same invariance. This can possibly rule out certain closures that
would violate such invariance, and point towards systematic procedures
to guarantee it. Here, the focus will be on invariance properties related to thermodynamics.\\

The remainder of this paper is organized as follows : section 2 asks
by how much the Gibbs function defining the thermodynamics of a multi-component
fluid may vary while leaving the observable predictions of the Navier-Stokes-Fourier
model unchanged. As expected, it is found that partial enthalpies
and entropies are defined up to arbitrary reference values, with only
the restriction that the reference values for species that may transform
into each other, such as the various phases of water, must vary together.
Section 3 focuses on conservative variables, whose definition involves
a reference pressure and which are often at the heart of predictive
models, as prognostic variables and as quantities involved in closure
relationships. Specifically, the dependence of commonly used conservative
variables on the reference pressure and on admissible changes of the
Gibbs function is examined. These results are used in section 4 to
examine the invariance of down-gradient closures. It is found that
flux-gradient closure relationships expressed in terms of conservative
variables lead generally to non-invariant models. A construction of
``reduced'' fluxes and gradients of entropy and enthalpy is proposed,
which can be used to construct closures that are invariant by design.
Section 5 summarizes the results and discusses their wider implications.

\section{Thermodynamic invariance of the Navier-Stokes-Fourier model}

\subsection{The Gibbs thermodynamic potential}

We consider a multicomponent fluid. A finite fluid parcel contains
masses $m_{a}$ of each constituent $a$, where $a$ belongs to a
finite set of symbolic values. For moist air containing various phases
of water not assumed to be at phase equilibrium, $a\in\{d,v,l,i\}$
for dry air, water vapor, liquid water and ice. For seawater, $a\in\{w,\sigma\}$
for water and salt. The masses sum up to the total mass of the fluid
parcel $m=\sum_{a}m_{a}$.

Important properties of the fluid parcel are its pressure $p$, temperature
$T$, volume $V$, entropy $S$, internal energy $U,$ enthalpy $H=U+pV$
and Gibbs free energy $G=H-TS$, among which $V,\,S,\,U,\,H,\,G$
are extensive. In the sequel we regard any extensive quantity $F$
as a function of $p,T,m_{a}$ (meaning that it depends on all masses
$m_{a}$), note $F^{p},F^{T}$ its derivatives with respect to $p,T$,
and 
\begin{equation}
f^{a}\equiv\frac{\partial}{\partial m_{a}}F(p,T,m_{a})
\end{equation}
the partial functions $f^{a}$ (in $J/kg$ if $F$ is an energy in
$J$). Especially important partial functions are partial enthalpies
$h^{a},$ partial entropies $s^{a}$ and chemical potentials $g^{a}=h^{a}-Ts^{a}$.
The extensive character of $H,\,S,\dots$ then implies :
\begin{equation}
H=h^{a}m_{a},\qquad S=s^{a}m_{a}\label{eq:partial_function}
\end{equation}
where Einstein's notation on repeated indices is used. It is to be
noted that (\ref{eq:partial_function}) is \emph{not} a definition
of $H,\,S$ from the partial enthalpies and entropies. Instead, it
\emph{follows} from the definition of $h^{a}$ and $s^{a}$ as derivatives
of $H,\,S$ and allows to \emph{interpret them} as the enthaply and
entropy attached to each component of the fluid. 

For applications to fluid flow, it is more convenient to consider
extensive functions per unit volume or per unit mass. For an extensive
function $F(p,T,m_{a})$, $F/V$ is obtained simply by evaluating
$F(p,T,\rho_{a})$ with $\rho_{a}\equiv m_{a}/V$ mass per unit volume
for species $a$. $F$ per unit mass $f=F/m$ is similarly obtained
as $f=F(p,T,q_{a})$ with $q_{a}$ the mass ratio $q_{a}=m_{a}/m$.
Since $\sum_{a}q_{a}=1$, the resulting $f(p,T,q_{a})$ is regarded
as a function of all mass ratios except one, usually chosen to be
$q_{d}=1-\left(q_{v}+q_{l}+q_{i}\right)$ for moist air, and $q_{w}=1-q_{\sigma}$
for seawater.

It follows from the first and second laws that :
\begin{align}
\text{d}G & =V\text{d}p-S\text{d}T+g^{a}\text{d}m_{a}.\label{eq:dG}
\end{align}
Especially :
\begin{equation}
V=G^{p},\quad S=-G^{T},\qquad S^{p}=-V^{T}\qquad H=G-TG^{T},\quad C_{p}=H^{T}=-TG^{TT}=TS^{T}\label{eq:thermo_identities}
\end{equation}
with $C_{p}$ heat capacity (in $J/K$) at constant pressure. More
generally, a fundamental property of $G$ is that it is a thermodynamic
potential, i.e. \emph{all thermodynamic functions can be expressed
from $G(p,T,m^{a})$, its derivatives and $p,T,m^{a}$} \citep{groot_non-equilibrium_1962,hauf_entropy_1987,thuburn_use_2017}.
Furthermore, $G$ has the unique property that its canonical variables
$p,T,m_{a}$ are directly measurable. We shall use this property as
follows. Consider:
\[
G(p,T,m_{a})=G^{\star}(p,T,m_{a})+\epsilon\,\delta G(p,T,m_{a})
\]
with $G^{\star}$ a ``base'' Gibbs function and $\epsilon\ll1$.
Then any thermodynamic function $X$ is of the form : 
\[
X(p,T,m_{a})=X^{\star}(p,T,m_{a})+\epsilon\,\delta X(p,T,m_{a})+O(\epsilon^{2}).
\]
A function $X$ such that $\delta X=0$ is said to be invariant. By
design $p,T,\rho_{a}$ are invariant.

We now seek to identify those variations $\delta G(p,T,m_{a})$ that,
given an initial state of the fluid given in terms of the observable
quantities $p,T,\rho_{a}$, leaves unchanged their future values predicted
by the NSF model. Mathematically, we demand that $\delta G$ be such
that $\partial/\partial t\,(\delta p,\delta T,\delta\rho_{a})=0$.
To identify the possible $\delta G$, we first write down the NSF
model in full.

\subsection{Navier-Stokes-Fourier equations}

We follow the standard approach of irreversible thermodynamics \citep{groot_non-equilibrium_1962}.
The conservation of mass implies :
\begin{equation}
\frac{\partial\rho_{a}}{\partial t}+\nabla\cdot\mathbf{F}_{a}=P_{a}\label{eq:irr_mas-1}
\end{equation}
with $P_{a}$ a production/destruction rate per unit volume due to
phase change / chemical reactions and $\mathbf{F}_{a}$ a mass flux.
By conservation of mass, $\sum_{a}P_{a}=0$. From mass fluxes, the
fluid velocity is defined as a mass-weighted barycentric velocity
:
\begin{equation}
\mathbf{u}\equiv\frac{1}{\rho}\sum_{a}\mathbf{F}_{a}
\end{equation}
with $\rho\equiv\sum_{a}\rho_{a},$ after which mass fluxes are decomposed
into an advective flux and a diffusive flux :
\begin{equation}
\frac{\partial\rho_{a}}{\partial t}+\nabla\cdot\left(\rho_{a}\mathbf{u}+\mathbf{j}_{a}\right)=P_{a}\label{eq:irr_mas-1-1}
\end{equation}
By design $\sum_{a}\mathbf{j}_{a}=0$ and :
\begin{equation}
\frac{\partial\rho}{\partial t}+\nabla\cdot\left(\rho\mathbf{u}\right)=0\label{eq:irr_mas-1-1-1}
\end{equation}

Similarly for entropy : 
\begin{equation}
\frac{\partial S}{\partial t}+\nabla\cdot\left(S\mathbf{u}+s^{a}\mathbf{j}_{a}+\mathbf{j'}_{s}\right)=P_{s}\label{eq:irr_entropy-1}
\end{equation}
with $S$ entropy per unit volume, $P_{s}\ge0$ a positive rate of
entropy production per unit volume, $\mathbf{j}_{s}=s^{a}\mathbf{j}_{a}+\mathbf{j'}_{s}$
an entropy flux split into a part $s^{a}\mathbf{j}_{a}$ due to diffusion
and a reduced entropy flux $\mathbf{j}'_{s}$ due to heat conduction.
Using $V=G^{p}$, the equation of state can be expressed as :
\begin{equation}
1=\frac{\partial}{\partial p}G(p,T,\rho_{a})\label{eq:EOS-1}
\end{equation}
regarded as an equation to be solved for pressure $p$. Momentum balance
includes pressure, gravity and viscosity : 
\begin{equation}
\frac{D\mathbf{u}}{Dt}+\frac{1}{\rho}\nabla p=-\nabla\Phi+\frac{1}{\rho}\nabla\cdot\boldsymbol{\tau}\label{eq:irr_viscosity-1}
\end{equation}
where $\Phi(\mathbf{x})$ is the geopotential field and we omit the
Coriolis force for simplicity.

Total energy conservation is achieved by letting the entropy production $P_{s}$
be :
\begin{equation}
TP_{s}=\varepsilon-\mathbf{j}'_{s}\cdot\nabla T-\mathbf{j}_{a}\cdot\nabla_{T}g^{a}-P_{a}g^{a}.\label{eq:irr_entropy_prod-1}
\end{equation}
with $\varepsilon$ the viscous rate of kinetic energy dissipation
and $\nabla_{T}g^{a}$ the isothermal gradient of chemical potential
:
\begin{equation}
\varepsilon=\nabla\mathbf{u}:\boldsymbol{\tau},\qquad\nabla_{T}g^{a}\equiv\nabla g^{a}+s^{a}\nabla T.\label{eq:irr_entropy_prod_invariant-1}
\end{equation}
The term $P_{a}g^{a}$ represents entropy production due to phase
change. Due to conservation of mass, it depends in fact on differences
between chemical potentials $g^{a}$ of species $a$ involved in phase
change. 

From these budgets, one derives the total energy budget :
\begin{equation}
\frac{\partial E}{\partial t}+\nabla\cdot\left(\mathbf{F}_{E}+\mathbf{j}_{E}\right)=0,\label{eq:irr_energy_budget-1}
\end{equation}
\begin{equation}
\text{where }E=\rho\frac{\mathbf{u}^{2}}{2}+\rho\Phi+H(p,s,\rho_{a})-p,\qquad\mathbf{F}_{E}=\left(\rho\frac{\mathbf{u}^{2}}{2}+\rho\Phi+H(p,s,\rho_{a})\right)\mathbf{u},\quad\mathbf{j}_{E}=\mathbf{j}_{H}+h^{a}\mathbf{j}_{a}-\mathbf{u}\cdot\boldsymbol{\tau}.
\end{equation}
with $\mathbf{j}_{H}\equiv T\mathbf{j}'_{s}$ the sensible heat flux
due to heat conduction.

\subsection{Admissible variations of the Gibbs function}

We now let $\delta(p,T,\rho_{a})=0$ and demand that $\delta\,(\partial p/\partial t,\,\partial T/\partial t,\,\partial\rho_{a}/\partial t)=0$.
Since $p$ is obtained by solving (\ref{eq:EOS-1}) for $p$:
\begin{equation}
\delta G^{p}=0
\end{equation}
Thus $\delta G=\delta G(T,m_{a})$ must be independent of $p$. Let
us first consider flow conditions such that, in a certain region,
$p,T,\rho_{a}$ are uniform and $\mathbf{\mathbf{u}}=0$. This region
of space behaves as a motionless fluid parcel of constant volume undergoing
phase change. Now that $V=G^{p}$ is invariant, the Helmholtz free
energy $F(V,T,m_{a})=G-pV$ satisfies $\delta F=\delta G$. $F$ is
useful because the heat capacity at constant volume $C_{v}=-T\partial_{TT}F$
and because temperature evolves while keeping internal energy and
volume constant :
\begin{equation}
0=C_{v}\frac{\partial T}{\partial t}+u^{a}P_{a}
\end{equation}
where $u^{a}=\partial_{a}U(V,T,m_{a})$. $C_{v}$ being experimentally
observable, both $C_{v}$ and $u^{a}P_{a}$ must be invariant so that
$\delta\partial_{TT}F=0$, i.e. $\delta F=\delta G$ is linear in
temperature :
\begin{equation}
\delta G=\delta H(m_{a})-T\delta S(m_{a}).
\end{equation}
with $\delta H$, $\delta S$ extensive functions of composition only.
This family spans the null space of the two equivalent problems of
(i) finding $G$ given $V(p,T,m_{a})=G^{p}$ and $C_{p}(p,T,m_{a})=-TG^{TT},$
and (ii) finding $F(V,T,m_{a})$ given $p(V,T,m_{a})=\partial_{V}F$
and $C_{V}(V,T,m_{a})=-T\partial_{TT}F$. 

We leave temporarily aside the contribution from phase change and
return to the NSF for a general initial flow configuration. The condition
$\delta\partial\rho_{a}/\partial t=0$ is in fact a condition on the
closures yielding $P_{a}$ and the diffusive flux $\mathbf{j}_{a}$.
Since $P_{a}$ are rates of condensation/evaporation/sublimation in
a fluid parcel, they must be observable given $p,T,\rho_{a}$ , hence
$\delta P_{a}=0$. Similarly if the closure for $\mathbf{j}_{a}$
guarantees that $\delta\mathbf{j}_{a}=0$, then $\delta\partial\rho_{a}/\partial t=0$.
It remains to determine under which condition for $\delta G$ the
predicted temperature is invariant, i.e. $\delta(\partial T/\partial t)=0$.
There are several possible, equivalent ways : (i) derive an evolution
equation for temperature and establish the conditions for its invariance
(ii) establish the conditions for the invariance of (\ref{eq:irr_entropy-1})
(iii) establish the conditions for the invariance of (\ref{eq:irr_energy_budget-1}).
In view of section 4, we choose here to examine the invariance of
(\ref{eq:irr_entropy-1}). 

(\ref{eq:irr_entropy-1}) is a particular form of the second law which
dictates how entropy varies. Although entropy itself may not be invariant,
the second law implies that entropy production $P_{s}$ is invariant.
Furthermore, since $P_{s}$ is a sum of contribution of independent
processes, we demand that each contribution be invariant separately.
The viscous and conductive contributions $\nabla\mathbf{u}:\boldsymbol{\tau},\,\mathbf{j}'_{s}\cdot\nabla T$
are invariant if the viscous stress tensor $\boldsymbol{\tau}$ and
the reduced entropy flux, or equivalently the sensible heat flux $\mathbf{j}_{H}=T\mathbf{j}'_{s}$
are invariant. This is again a requirement for the respective closures.
Similarly, the contribution from diffusion $\mathbf{j}_{a}\cdot\nabla_{T}g^{a}$
must be invariant, which will be the case if : 
\begin{equation}
0=\delta\nabla_{T}g^{a}=\delta\nabla h^{a}-T\delta\nabla s^{a}.
\end{equation}
where $\delta h^{a}$ and $\delta s^{a}$ are functions of composition
only. Thus, $\delta\nabla_{T}g^{a}=0$ is achieved for arbitrary fields
$T(\mathbf{x}),\,\rho_{a}(\mathbf{x})$ if $\delta h^{a}$ and $\delta s^{a}$
are constants. This corresponds to the statement that partial enthalpies
and entropies are determined up to an arbitrary constant. To finally
examine the contribution from phase change $P_{a}g^{a}$, let us consider
first the simple case of condensation of water vapor into liquid water
:
\begin{equation}
P_{s}=-C,\qquad P_{l}=C,\qquad TP_{s}=\cdots+C(g^{l}-g^{w})
\end{equation}
Thus $\delta P_{s}=0$ imposes that $\delta\left(g^{l}-g^{w}\right)=0$.
Since $\delta g^{a}=\delta h^{a}-T\delta s^{a}$ with constant $\delta h^{a}$
and $\delta s^{a}$, this requires $\delta(h^{l}-h^{w})=0.$ This
condition results from the fact that $h^{l}-h^{w}$ is the latent
heat of evaporation of water, an observable quantity which must be
invariant. Similarly the condition for phase equilibrium is the equality
of chemical potentials $g^{l}=g^{w}$, which must be invariant. 

Using $S=\rho_{a}s^{a},\,\delta S=\rho_{a}\delta s^{a}$, it is finally
easy to check that the remainder of the entropy budget is invariant
without further conditions on $\delta G$:
\begin{equation}
\delta\left(\frac{\partial S}{\partial t}+\nabla\cdot\left(S\mathbf{u}+s^{a}\mathbf{j}_{a}\right)\right)=\delta s^{a}\left(\frac{\partial}{\partial t}\rho_{a}+\nabla\cdot\left(\rho_{a}\mathbf{u}+\mathbf{j}_{a}\right)\right)=0
\end{equation}
\\

To conclude, the predictions of the Navier-Stokes-Fourier model are
insensitive to variations $\delta G$ of the Gibbs function of the
form 
\begin{equation}
\delta G=m_{a}\delta h^{a}-Tm_{a}\delta s^{a}\label{eq:dG-1}
\end{equation}
where $\delta h^{a}$ and $\delta s^{a}$ are constants whose arbitrariness
is limited only by the conditions $\delta h^{a}=\delta h^{b}$ and
$\delta s^{a}=\delta s^{b}$ for species $a,b$ that transform into
each other via phase change. 

\section{Invariance of conservative variables}

In theoretical and numerical models of atmospheric and oceanic motion,
separate mass budgets for each species/phase and for entropy are usually
replaced by an equivalent formulation including conservation of total
fluid density $\rho=\sum_{a}\rho_{a}$ and transport of composition
and specific entropy :
\begin{equation}
\frac{\partial\rho}{\partial t}+\nabla\cdot\rho\mathbf{u}=0,\qquad\rho\frac{D}{Dt}q_{a}=P_{a},\qquad\rho\frac{Ds}{Dt}=P_{s}
\end{equation}
where $D/Dt$ is the material (Lagrangian) derivative. Furthermore
it is often preferred to prognose, instead of specific entropy, a
conservative variable $\theta$. Here, we adopt the perhaps narrow
definition that $\theta$ is an entropic variable, i.e. a given function
of specific entropy and composition \citep{hauf_entropy_1987,eldred_thermodynamically_2022}.
This function usually involves conventional constants, e.g. a reference
pressure $p_{0}$:
\begin{equation}
\theta\equiv\theta(s,q^{a},p_{0}).
\end{equation}
This definition ensures that $D\theta/Dt=0$ in adiabatic conditions.
We examine how usual conservative variables depend on reference
enthalpies, entropies and pressure. Introducing the reference pressure
$p_{0}$ potentially introduces a new contribution, proportional to
$\delta p_{0}$, to the variations $\delta F$ of a thermodynamic
function $F$ at fixed observable quantities. 

\subsection{Potential temperature}

Potential temperature is defined, for arbitrary thermodynamics, as the temperature a
fluid parcel would have after evolving adiabatically from conditions
$(p,T,q_{a})$ to conditions $(p_{0},\theta,q_{a})$. It is therefore
defined as the solution of :
\begin{equation}
s(p,T,q_{a})=s(p_{0},\theta,q_{a}).
\end{equation}
This definition applies to moist air in which water phases are assumed
at equilibrium, $\theta$ being then liquid potential temperature
\citep{hauf_entropy_1987}. The exact expression of $\theta_{l}$
is often replaced by approximate expressions of various accuracies
\citep{betts_empirical_1973,marquet_definition_2011}. 

On the l.h.s, $\delta s=q_{a}\delta s^{a}$ while on the r.h.s $\delta s=q_{a}\delta s^{a}+s^{T}(\theta,p_{0},q_{a})\delta\theta+s^{p}(\theta,p_{0},q_{a})\delta p_{0}$.
Thus $\theta$ depends on $p_{0}$, but not on reference entropies,
which is expected since temperature is observable. Using (\ref{eq:thermo_identities})
yields :
\begin{equation}
\frac{\delta\theta}{\theta}=\frac{v^{T}}{c_{p}}\delta p_{0}\label{eq:dtheta_general}
\end{equation}
where $v(p,T,q_{a})$ is specific volume and $v^{T}/c_{p}$ is evaluated
at $(p_{0},\theta,q_{a})$. For unsaturated air modelled as an ideal
perfect gas, $\theta$ and $\delta\theta$ takes the simple expression
:
\begin{equation}
\theta=T\left(\frac{p}{p_{0}}\right)^{-R/c_{p}}
\end{equation}
where $R$ and $c_{p}$ depend on specific humidity, so that :
\begin{equation}
\frac{\delta\theta}{\theta}=\frac{R}{c_{p}}\frac{\delta p_{0}}{p_{0}}.\label{eq:dtheta_perfectgas}
\end{equation}
\citep{lebonnois_superrotation_2010} introduces the idea of modelling
``air'' on Venus as a perfect gas, albeit non ideal due to a temperature
dependent heat capacity $c_{p}=c_{p}(T)$. Still, $pv=RT$ implies
$v^{T}=R/p$ so that (\ref{eq:dtheta_perfectgas}) holds in this case
with $c_{p}=c_{p}(\theta)$. 

Moist air is often modelled by (i) neglecting the volume of the condensed
phases (ii) modelling the gaseous phase as an ideal mixture of dry
air and water vapor \citep{lauritzen_ncar_2018}. Total mass is $m=m_{g}+m_{c}$
with $m_{g}=m_{d}+m_{v}$ the gaseous mass and $m_{c}=m_{l}+m_{i}$
the condensate mass. With hypotheses (i-ii), $pV=mRT$ holds where
$R(q_{a})=(m_{d}R^{d}+m_{v}R^{v})/m$ takes into account all water
phases. Specific volume is $v=V/m=RT/p$ so that (\ref{eq:dtheta_perfectgas})
still holds. 

In all cases $\delta\theta=\beta\delta p_{0}$ depends on $\delta p_{0}$
by a factor $\beta(p_{0},\theta,q_{a})$ depending on $\theta$ and
composition, directly or via other conservative variables. Except
for an ideal perfect gas, for which the ratio $R/C_{p}$ is constant,
$\beta$ is a nonlinear function of $\theta,q_{a}$. It will appear
in section 4 that this nonlinearity is the root cause of the violation
of invariance of down-gradient closures.

\subsection{Virtual potential temperature and potential volume}

Virtual temperature $T_{v}$ of moist air is defined as the temperature
a dry air parcel should take, at a given pressure, in order to match
the density of moist air, assuming perfect gases for dry and moist
air. Thus, up to the multiplicative constant $R^{d}$, virtual temperature
is equivalent to the product $pv$ with $v=V/m$ specific volume.

Virtual potential temperature $\theta_{v}$ is the virtual temperature
a moist air parcel would acquire after an adiabatic change of pressure
to a reference pressure $p_{0}$. Similarly, $\theta_{v}$ is equivalent,
up to a multiplicative constant, to potential specific volume, another entropic variable defined
as specific volume evaluated at $(p_{0},\theta,q_{a})$ instead of
$(p,T,q_{a})$ :
\begin{equation}
v_{pot}=v(p_{0},\theta,q_{a}).
\end{equation}
Using $\text{d}(pv)=\left(pv\right)^{p}\text{d}p+(pv)^{T}\text{d}T+\dots$
one gets :
\begin{equation}
\delta(p_{0}v_{pot})=\left(pv\right)^{p}\delta p_{0}+(pv)^{T}\delta\theta
\end{equation}
where $\left(pv\right)^{p,T}$ are derivatives of the function $p\,v(p,T,q_{a})$
evaluated at $(p_{0},\theta,q_{a})$. With a gaseous part assumed
to be a perfect gas, and with the condensate volume being neglected,
$pv$ is a function of $T$ and $q_{a}$, linear in $T$. With these
assumptions $\left(pv\right)^{p}=0$,~$(pv)^{T}=p_{0}v_{pot}/\theta$
and : 
\begin{equation}
\frac{\delta\left(p_{0}v_{pot}\right)}{p_{0}v_{pot}}=\frac{\delta\theta}{\theta}.\label{eq:dvpot}
\end{equation}
(\ref{eq:dvpot}) show that $p_{0}v_{pot}$, hence $\theta_{v}$,
is sensitive to $p_{0}$ very much like $\theta$, i.e. $\delta\theta_{v}=\beta_{v}(p_{0},\theta,q_{a})\delta p_{0}$
with $\beta_{v}$ a non-linear function of $\theta_{v},q_{a}$.

\subsection{Moist entropy potential temperature}

A recent addition to the list of moist conservative variables is moist
entropy potential temperature $\theta_{s}$, defined by (\citep{marquet_definition_2011},
eq. 39):
\begin{equation}
s=s_{r}+c_{p}^{d}\log\frac{\theta_{s}}{\theta_{sr}}
\end{equation}
where $\theta_{sr},\,s_{r}$ are constant reference values. Thus :
\begin{equation}
\delta\theta_{s}=\frac{\theta}{c_{p}^{d}}\left(q_{a}\delta s^{a}-\delta s_{r}\right)
\end{equation}
Thus $\theta_{s}$ is sensitive to a change of reference entropies,
and $\delta\theta_{s}$ is a nonlinear function of $\theta,q_{a}$.

\subsection{Conservative temperature and potential enthalpy}

In addition to potential temperature, which is sensitive to reference
pressure following (\ref{eq:dtheta_general}) with a complex seawater
equation of state \citep{feistel_thermodynamic_2010}, ocean modellers
use conservative temperature $T_{c}$ , which is equivalent to potential
enthalpy $h_{pot}(s,q_{a})$ up to additive and multiplicative constants
\citep{mcdougall_potential_2003}:
\begin{equation}
h(p_{0},s,q_{a})=h_{pot}=h_{0}+c_{p}^{0}T_{c}.
\end{equation}
Potential enthalpy is sensitive to reference pressure and reference
enthalpies :
\begin{equation}
\delta h_{pot}=\theta q_{a}\delta s_{a}+v_{pot}\delta p_{0}\label{eq:invariance_hpot}
\end{equation}
with $\theta,\,v_{pot}$ potential temperature and potential volume.
Conservative temperature is sensitive in addition to the choice of
the constants $h_{0}$ and $c_{p}^{0}$ :
\begin{equation}
c_{p}^{0}\delta T_{c}=\theta q_{a}\delta s_{a}+v_{pot}\delta p_{0}-\delta h_{0}-T_{c}\delta c_{p}^{0}.
\end{equation}
However these extra variations are affine in $T_{c}$, and thus harmless
for invariance, as shown in section 4.

(\ref{eq:invariance_hpot}) shows that $p_{0}$ affects $h_{pot}$
with a coefficient $v_{pot}$, which is a nonlinear functions of $h_{pot}$
and $q_{a}$ except if the equation of state $v(p,T_{c},q_{a})$ has
been linearized with respect to $T_{c},\,q_{a}$. Even so, variations
$\delta s_{a}$ contribute to $\delta h_{pot}$ with the coefficient
$\theta q_{a}$ which is a nonlinear function of $T_{c},\,q_{a}$.

\subsection{Buoyancy}

Potential temperature and virtual potential temperature are often
used as proxies for buoyancy : buoyancy-related quantities are approximately
evaluated using $\theta,\,\theta_{v}$, their profiles and their fluxes.
If the quantity to be evaluated is invariant, but its approximate
expression is not, invariance of the whole model is jeopardized. 

An important buoyancy-related quantity is the Brunt-Vaisala pulsation
$N$, potentially involved in turbulence closures through the Richardson
number $Ri$. For vertical profiles $p(z),\,T(z),\,q_{a}(z)$, with
$\rho(z)=(-\text{d}p/\text{d}z)/g$ in hydrostatic balance with $p(z)$,
the general expression of $N^{2}$ is :
\begin{equation}
N^{2}=\rho g\left[\frac{\text{d}v}{\text{d}z}-\frac{\text{d}p}{\text{d}z}\frac{\partial}{\partial p}v(p,s(z),q_{a}(z))\right]\label{eq:BV_exact-1}
\end{equation}
where the second term on the r.h.s discards vertical variations of
specific volume due to pressure, to keep only those due to vertical
variations of entropy and composition. Indeed one may replace $s$
in (\ref{eq:BV_exact-1}) by any entropic variable $\theta$ (which
could be $\theta_{v}$) and rearrange (\ref{eq:BV_exact-1}) as :
\begin{equation}
N^{2}=\rho g\left[\frac{\text{d}\theta}{\text{d}z}\,\frac{\partial}{\partial\theta}v\left(p,\theta,q_{a}\right)+\sum_{a}\frac{\text{d}q_{a}}{\text{d}z}\,\frac{\partial}{\partial q_{a}}v\left(p,\theta,q_{a}\right)\right]\label{eq:BV_exact}
\end{equation}
While (\ref{eq:BV_exact}) is more useful, (\ref{eq:BV_exact-1})
makes it manifest that $N^{2}$ is invariant, as it should since $N^{2}$
is related to the pulsation of internal gravity waves, which are observable.
From (\ref{eq:BV_exact}) it is clear that, except by chance, $N^{2}$
is not exactly proportional to the vertical gradient of a single entropic
variable, but involves also the vertical gradient of composition.
Furthermore the coefficient $\partial v/\partial q_{a}$ is a derivative
of specific volume at constant $p$ and $\theta$, which is not invariant.
Thus $\partial v/\partial q_{a}$ is not invariant and letting for
instance $N^{2}\simeq(g/\theta_{v})\text{d}\theta_{v}/\text{d}z$
is not only an approximation, but also breaks the invariance of $N^{2}$.

Similarly, the turbulent kinetic energy (TKE) budget includes a production
term due to buoyancy forces, often called the ``buoyancy flux''\footnote{It is expected from fluxes that their divergence is involved in the
flux-form budget of a related quantity. This is not the case of the
``buoyancy flux'', which appears only in the TKE budget as a source
term, hence the quotes. }. Since TKE is invariant, so is the buoyant production of TKE. In
an anelastic context, the buoyancy force is :
\begin{equation}
b=g\left(\rho_{ref}v(p_{ref},\theta,q_{a})-1\right)\label{eq:buoyancy}
\end{equation}
with $p_{ref}(z)$ the reference pressure profile involved in the
definition of the anelastic approximation, $\rho_{ref}=(-\text{d}p_{ref}/\text{d}z)/g$
the corresponding density profile, and $\theta$ any entropic variable
\citep{pauluis_thermodynamic_2008}. Thus the buoyant production is
:
\begin{equation}
\overline{b'w'}=g\rho_{ref}\overline{v'w'}\label{eq:buoyancy_flux}
\end{equation}
where $v'$ are fluctuations of specific volume at the fixed pressure
$p_{ref}(z)$ due to fluctuations of $\theta$ and $q_{a}$. Since
$b$ is an invariant quantity, so is (\ref{eq:buoyancy_flux}). We
may linearize fluctuations around mean values $\overline{\theta},\,\overline{q_{a}}$,
leading to :
\begin{equation}
\overline{b'w'}=g\rho_{ref}\left[\overline{\theta'w'}\frac{\partial}{\partial\theta}v\left(p_{ref},\overline{\theta},\overline{q_{a}}\right)+\sum_{a}\overline{q_{a}w'}\,\frac{\partial}{\partial q_{a}}v\left(p_{ref},\overline{\theta},\overline{q_{a}}\right)\right].\label{eq:buoyancy_flux-1}
\end{equation}
(\ref{eq:buoyancy_flux-1}) approximates (\ref{eq:buoyancy_flux})
and is similar in structure to (\ref{eq:BV_exact}). Therefore, this
expression of $\overline{b'w'}$ is also invariant. As in (\ref{eq:BV_exact}),
invariance results from the contribution of composition. If an approximate
expression involving only the flux of an entropic variable is used,
such as $\overline{b'w'}\simeq(g/\theta_{v})\overline{\theta'_{v}w'}$,
invariance is jeopardized again.

\section{Invariance of down-gradient closures}

Section 3 has shown a non-trivial dependence of conservative variables
to reference values. It is thus natural to ask whether standard closure
assumptions such as down-gradient fluxes lead to invariant models.
This question is asked first for closures applied to potential temperature,
in which case the answer is, in general, negative. Thus we develop
an alternative approach whereby invariant ``reduced gradients''
and ``reduced fluxes'' are constructed, so that they can serve as
the basis for closure relationships. Comparing closures based on potential
temperature to such invariant closures reveals the role of usually
neglected source terms.

\subsection{Down-gradient closure for potential temperature}

It is assumed that potential temperature $\theta$ has been decomposed
into an average $\overline{\theta}$ and fluctuations $\theta'$,
with the goal of modelling averages \citep{mellor_development_1982}.
Furthermore, following current practice \citep{lauritzen_reconciling_2022},
we assume that the same thermodynamic relationships as before are
used between averages $\overline{\theta},\,\overline{T},\,\overline{q_{a}},\,\dots$.
Then $\overline{\theta}$ and its gradient are sensitive to $p_{\text{0}}$
following:
\begin{equation}
\delta\overline{\theta}=\beta(p_{0},\overline{\theta},\overline{q_{a}})\delta p_{0},\qquad\delta\frac{\partial\overline{\theta}}{\partial z}=\left(\frac{\partial\beta}{\partial\theta}\frac{\partial\overline{\theta}}{\partial z}+\frac{\partial\beta}{\partial q_{a}}\frac{\partial\overline{q_{a}}}{\partial z}\right)\delta p_{0}
\end{equation}
Averaging applied to the adiabatic transport equation for $\theta$
and $q_{a}$ yields :
\begin{align}
\frac{\partial}{\partial t}\rho\overline{\theta}+\frac{\partial}{\partial z}\rho\overline{\theta'w'} & =0,\label{eq:turbulent_budget_theta}\\
\frac{\partial}{\partial t}\rho\overline{q_{a}}+\frac{\partial}{\partial z}\rho\overline{q_{a}'w'} & =0,
\end{align}
with $\rho\overline{\theta'w'}$ and $\rho\overline{q_{a}'w'}$ vertical
turbulent fluxes and we assume horizontal statistical homogeneity
for simplicity. As seen in (\ref{eq:irr_energy_budget-1}), although
the non-invariance of $\theta$ propagates to its gradient, a certain
choice of closure for $\rho\overline{\theta'w'}$ could conceivably
result in an invariant predictive model. As a tractable example, we
consider a down-gradient closure for the turbulent fluxes :
\begin{align}
\frac{\partial}{\partial t}\rho\overline{\theta}-\frac{\partial}{\partial z}\rho K_{z}\frac{\partial\overline{\theta}}{\partial z} & =0.\label{eq:down_gradient}\\
\frac{\partial}{\partial t}\rho\overline{q_{v}}-\frac{\partial}{\partial z}\rho K_{z}\frac{\partial\overline{q_{v}}}{\partial z} & =0.\label{eq:down_gradient_q}
\end{align}
where $K_{z}$ is also to be determined, typically from a mixing length,
a Richardson number and other quantities, all invariant \citep{mellor_development_1982}.
To determine whether (\ref{eq:down_gradient}-\ref{eq:down_gradient_q})
is invariant, we apply $\delta$ to (\ref{eq:down_gradient}) and
find :
\begin{align}
\delta\left(\frac{\partial}{\partial t}\rho\overline{\theta}-\frac{\partial}{\partial z}\rho K_{z}\frac{\partial\overline{\theta}}{\partial z}\right) & =\left(\frac{\partial\beta}{\partial t}-\frac{\partial}{\partial z}\rho K_{z}\frac{\partial\beta}{\partial z}\right)\delta p_{0}\label{eq:invariance_downgradient_theta}
\end{align}
Since $\overline{\theta},\overline{\,q_{a}}$ obey an advection-diffusion
equation with the same turbulent diffusivity, $\beta=\beta(p_{0},\overline{\theta},\overline{\,q_{a}})$
obeys a similar equation \emph{albeit with a source term}:
\begin{equation}
\frac{\partial\beta}{\partial t}-\frac{\partial}{\partial z}\rho K_{z}\frac{\partial\beta}{\partial z}=\rho K_{z}\left(\frac{\partial^{2}\beta}{\partial\theta^{2}}\left(\frac{\partial\overline{\theta}}{\partial z}\right)^{2}+2\sum_{a}\frac{\partial^{2}\beta}{\partial\theta\partial q_{a}}\frac{\partial\overline{\theta}}{\partial z}\frac{\partial\overline{q_{a}}}{\partial z}+\sum_{a,b}\frac{\partial^{2}\beta}{\partial q_{a}\partial q_{b}}\frac{\partial\overline{q}_{a}}{\partial z}\frac{\partial\overline{q}_{b}}{\partial z}\right).
\end{equation}

If thermodynamic assumptions imply that $\beta$ depends linearly
on $\theta,q_{a}$, we can conclude that a down-gradient closure for
$\overline{w'\theta'}$ and $\overline{w'q_{a}'}$, with the same
turbulent diffusion coefficient, leads to a model prediction independent
from $p_{0}$. Otherwise the r.h.s of (\ref{eq:invariance_downgradient_theta})
does not vanish identically, i.e. invariance is violated by the model.
This is the case for all conservative quantities considered in section
3 except potential temperature of dry air. Thus a different, systematic strategy to guarantee
the invariance of models including down-gradient closures is now developed.

\subsection{Reduced gradients and fluxes}

In section 2, the energy and entropy budgets are invariant despite
the non-invariance of individual terms. The reason is twofold. Firstly,
because entropy and enthalpy depend in a simple, linear way on reference
enthalpies and entropies :
\begin{equation}
\delta S_{a}=m_{a}\delta s^{a}\quad\Rightarrow\delta\left(\frac{\partial S}{\partial t}+\nabla\cdot\left(S\mathbf{u}+s^{a}\mathbf{j}_{a}\right)\right)=0
\end{equation}
Secondly, this identity leads to the decomposition of the entropy
flux as:
\begin{equation}
\mathbf{j}_{s}=s^{a}\mathbf{j}_{a}+\mathbf{j'}_{s}
\end{equation}
with $\mathbf{j'}_{s}$ invariant. We can apply a similar decompositionto
the turbulent fluxes of entropy and enthalpy : 
\begin{equation}
\mathbf{J}'_{s}=\rho\overline{s'\mathbf{u}'}-s^{a}\mathbf{J}{}_{a},\qquad\mathbf{J}'_{H}=\rho\overline{h'\mathbf{u}'}-h^{a}\mathbf{J}{}_{a}\qquad\text{with }\mathbf{J}{}_{a}\equiv\rho\overline{q'_{a}\mathbf{u}'}\label{eq:reduced_flux}
\end{equation}
where $\mathbf{J}'_{s}$ is a reduced turbulent entropy flux and $\mathbf{J}'_{H}$
the turbulent sensible heat flux. Similarly, since the gradient of specific
entropy depends linearly on reference entropies, substracting from $\nabla s$ its non-invariant part yields a ``reduced
gradient'' $\widetilde{\nabla}s$ :
\begin{equation}
\delta\nabla s=\delta s^{a}\nabla q_{a}, \quad
\widetilde{\nabla}s\equiv\nabla s-s^{a}\nabla q_{a}\qquad
\Rightarrow\qquad \delta\widetilde{\nabla}s=0.\label{eq:reduced_gradient}
\end{equation}
which is invariant, as is also manifest in expression:
\begin{equation}
\widetilde{\nabla}s=\frac{c_{p}}{T}\nabla T-v^{T}\nabla p.
\end{equation}
obtained using (\ref{eq:thermo_identities}).
Thus, an invariant-by-design model can be obtained by prognosing mean
entropy by an averaged entropy budget including a closure relationship
between $\mathbf{J'}_{s}$ and known invariant quantities, such as
$\widetilde{\nabla}s$ and $\nabla q_{a}$.

\subsection{Invariant down-gradient closures}

Upon averaging, the mass and entropy budgets become 
\begin{equation}
\rho\left(\frac{\partial}{\partial t}+\overline{\mathbf{u}}\cdot\nabla\right)\overline{q_{a}}+\nabla\cdot\mathbf{J}_{a}=\overline{P_{a}},\qquad\rho\left(\frac{\partial}{\partial t}+\overline{\mathbf{u}}\cdot\nabla\right)\overline{s}+\nabla\cdot\mathbf{J}_{s}=\overline{P_{s}}.\label{eq:average_budgets}
\end{equation}
where closure formulae are needed for $\mathbf{J}{}_{s},\,\mathbf{J}_{a}$
(and also $\overline{P_{s}},\,\overline{P_{a}}$) in order to obtain
a predictive model. As previously explained, instead of searching
directly a closure for $\mathbf{J}{}_{s},\,\mathbf{J}_{a}$ in terms
of $\nabla\overline{s},\,\nabla\overline{q_{a}}$, we rewrite the
entropy budget as : 
\begin{equation}
\rho\left(\frac{\partial}{\partial t}+\overline{\mathbf{u}}\cdot\nabla\right)\overline{s}+\nabla\cdot\left(s^{a}\mathbf{J}_{a}\right)+\nabla\cdot\mathbf{J}'_{s}=\overline{P_{s}}.
\end{equation}
Now, we need closure formulae for $\mathbf{J}'_{s},\,\mathbf{J}_{a}$.
Since these fluxes are invariant, a closure taking as inputs $\widetilde{\nabla}\overline{s},\,\nabla\overline{q_{a}}$
will be invariant too. Thus, independently from the accuracy of the
closure, an invariant model is obtained. Let us examine the simple
case of down-gradient closures with a single turbulent diffusivity
$K$ :
\begin{equation}
\mathbf{J}_{a}=-\rho K\nabla q_{a},\qquad\mathbf{J}_{s}'=-\rho K\widetilde{\nabla}s.\label{eq:invariant_closure_q}
\end{equation}
Then
\begin{equation}
\mathbf{J}{}_{s}=-\rho K\left(\nabla s-s^{a}\nabla q_{a}\right)-\rho Ks^{a}\nabla q_{a}=-\rho K\nabla s\label{eq:invariant_closure_s}
\end{equation}
i.e. a down-gradient closure for the full entropy flux $\mathbf{J}_{s}$ is obtained. At
contrast to down-gradient closures for conservative variables, we
know without further investigation that the resulting model (\ref{eq:average_budgets})
with (\ref{eq:invariant_closure_q}-\ref{eq:invariant_closure_s})
is invariant.

It is intriguing that a down-gradient closure for entropy yields an
invariant model while a down-gradient closure for an entropic variable
seems, in general, not to. To clarify this point, we derive from (\ref{eq:average_budgets})
the evolution equation for $\hat{\theta}=\theta(\overline{s},\overline{q_{a}})$
: 
\begin{equation}
\rho\left(\frac{\partial}{\partial t}+\overline{\mathbf{u}}\cdot\nabla\right)\hat{\theta}+\nabla\cdot\mathbf{J}_{\theta}=\widehat{P_{\theta}}\label{eq:averaged_budget_theta}
\end{equation}
with 
\begin{equation}
\mathbf{J}_{\theta}\equiv\frac{\partial\theta}{\partial s}\mathbf{J}_{s}+\frac{\partial\theta}{\partial q_{a}}\mathbf{J}_{a},\qquad\widehat{P_{\theta}}\equiv\frac{\partial\theta}{\partial s}\overline{P_{s}}+\frac{\partial\theta}{\partial q_{a}}\overline{P_{a}}+\mathbf{J}_{s}\cdot\nabla\frac{\partial\theta}{\partial s}+\mathbf{J}_{a}\cdot\nabla\frac{\partial\theta}{\partial q_{a}}\label{eq:closure_theta}
\end{equation}
a turbulent flux and a mean production term. Assuming the same down-gradient
closure as above for $\mathbf{J}_{s},\,\mathbf{J}_{a}$ yields indeed
a down-gradient closure for the turbulent flux of $\theta$ : 
\begin{equation}
\mathbf{J}_{\theta}=-\rho K\nabla\hat{\theta}\label{eq:closure_theta-1}
\end{equation}

Being derived from an invariant model, ((\ref{eq:averaged_budget_theta})-(\ref{eq:closure_theta-1}))
is invariant. The significant difference between (\ref{eq:averaged_budget_theta})
and (\ref{eq:turbulent_budget_theta}) is that (\ref{eq:averaged_budget_theta})
involves a source term, while (\ref{eq:turbulent_budget_theta})
does not. Furthermore this source term :
\begin{equation}
\widehat{P_{\theta}}\equiv\frac{\partial\theta}{\partial s}\overline{P_{s}}+\frac{\partial\theta}{\partial q_{a}}\overline{P_{a}}-\rho K\left(\nabla s\cdot\nabla\frac{\partial\theta}{\partial s}+\nabla q_{a}\cdot\nabla\frac{\partial\theta}{\partial q_{a}}\right)
\end{equation}
appears not to be invariant, in general. Thus, invariance of (\ref{eq:averaged_budget_theta})
results from a compensation between the variations of the l.h.s of
(\ref{eq:turbulent_budget_theta}) and those of $\widehat{P_{\theta}}$.
Neglecting $\widehat{P_{\theta}}$, which may be justified by its
smallness compared to $\nabla\cdot\mathbf{J}_{\theta}$, breaks the
invariance of the model.

\section{Conclusion}

It is common for physical theories to involve scalar or vector potentials
which can be subjected to certain transformations without affecting
the observable predictions of the theory. This work raises the issue
of the invariance of models of geophysical flow motion with respect
to changes of arbitrary constants appearing in thermodynamic potentials and conservative variables,
an issue that has been so far ignored or at least overlooked. \\

Section 2 examines the ``degrees of arbitrariness'' present in the
standard, non-averaged Navier-Stokes-Fourier model of fluid motion,
including diffusion, conduction and phase change. It is found that
the NSF model is invariant with respect to arbitrary changes of reference
partial enthalpies and entropies, with the restriction that constants
for species that can transform into each other through phase change
vary identically. It is easy to verify that the same invariance applies
to sound-proof models with a consistent treatment of thermodynamics
and irreversible processes \citep{klein_thermodynamic_2012,eldred_thermodynamically_2021}.

While expected, this result implies that no other restriction applies.
Especially, assuming specific values for reference partial entropies
and/or enthalpies of dry air and water vapor \citep{marquet_definition_2011,marquet_computation_2015}
effectively imposes $\delta s^{a}=0$ and/or $\delta h^{a}=0$ and
is an unwarranted restriction. Any observation in a flow that can
reasonably be modelled by the NSF model must be treated as invariant
with respect to changes of reference enthalpies and entropies, since
such changes have no observable effect in the NSF model. Admittedly the NSF model
is insufficient to fully describe cloud physics,
as this would require additional processes such as surface tension
and interaction with aerosols. But these processes involve additional forms of energy,
and do not provide additional constraints on partial enthalpies and entropies. 
Only the study of chemical reactions
consuming oxygen and producing water would constrain the differences between partial enthalpies and entropies
of the species involved.
But since such reactions are not occurring in clouds, these constraints do not apply to a hypothetical
model incorporating all the physical processes relevant for clouds.
Thus, empirical relationships drawn from observations but requiring $\delta s^{a}=0$
and/or $\delta h^{a}=0$ to be invariant can only be due to chance,
i.e. are spurious and accidental. It is especially the case of relationships
involving moist entropy potential temperature \citep{marquet_definition_2011}.\\

Section 3 has established how usual conservative variables depend
on their reference pressure and, for certain of them, on reference
enthalpies and entropies. While this dependence is not quite a discovery,
its implications for closures had not been examined so far, which
has been done in section 4. It is found that when down-gradient closure
relationships are applied to conservative variables, the invariance
of the resulting models is violated, except in restricted cases such
as a dry ideal perfect gas. The problem of systematically constructing
invariant models is addressed by requiring that closures be based
on invariant quantities only. Invariant expressions (\ref{eq:reduced_flux}-\ref{eq:reduced_gradient})
for ``reduced gradients'' and ``reduced fluxes'' are proposed,
leading to invariant-by-design down-gradient closures. Within the family of
down-gradient closures, it is permissible to let turbulent diffusivities for
entropy and composition differ, and even to allow for cross-diffusivity. Using a single turbulent
diffusivity for entropy and composition is equivalent to applying the same closure to any conservative variable,
and yields an invariant model provided a matching source term
is included. 

These source terms are often neglected, which violates invariance.
Such a neglect may be motivated by the smallness of the source term
compared to other terms. However a non-invariant term is not intrinsically
small : it can be made arbitrarily large by a proper choice of the
reference value it depends on. This finding also calls for a reexamination
of the widespread practice of neglecting such source terms in physical
and numerical models. Certainly they must not be neglected in physical
(in the sense of theoretical) models in order to maintain invariance
and permit further theoretical analysis. Perhaps it is nevertheless
acceptable to neglect them in numerical models, especially if the
conservative variable and reference values have been chosen to minimize
the magnitude of the neglected production term, which is the case
at least for seawater conservative temperature \citep{mcdougall_potential_2003}. 

Overall, these results call for a systematic examination of the invariance
of existing models and closures. They suggest that closures should,
as much as possible, be relationships between thermodynamically invariant
quantities. The present analysis is limited to simple, local, down-gradient closures, and
its extension to the more complex and non-local closures used to model
shallow \citep{rio_thermal_2008} and deep convection \citep{rochetin_deep_2014}
may require the introduction of new invariant quantities beyond the
reduced gradients and fluxes proposed in section 4. Also, several
authors have recently explored data-driven approaches to construct
closure relationships \citep{rasp_deep_2018,mooers_assessing_2021,wang_stable_2022}.
Such data-driven models will likely violate thermodynamic invariance,
unless they are built from closure problems involving only invariant
quantities. At the same time, for consistency with the first and second law, data-driven closures
should probably predict, among others, energy and/or entropy fluxes rather
than raw temperature tendencies. Predicting reduced fluxes \eqref{eq:reduced_flux} would 
make these two desirable properties compatible. Eventually, it should be possible to combine the accuracy
allowed by data-driven approaches with the consistency guaranteed
by the formulation of invariant closure problems. Such closures may
also have better generalization capabilities, a generic concern with
data-driven approaches.\\

Only the issue of invariance is dealt with here. Other questions,
such as the conditions for coarse-grain models with closures to conserve
total energy and increase total entropy, are left for future work. Preliminary
results suggests that the introduction of reduced fluxes and gradients
permits a quite general treatment of this question for down-gradient
closures.\\

This work is part of the AWACA project that has received funding from
the European Research Council (ERC) under the European Union’s Horizon
2020 research and innovation programme (Grant agreement No. 951596)




\bibliography{biblio}



\end{document}